\documentclass[twocolumn,showpacs,preprintnumbers,amsmath,amssymb]{revtex4}

\usepackage{graphicx,psfrag}
\usepackage{amssymb}
\usepackage{amsmath}
\usepackage[squaren,Gray,textstyle]{SIunits}
\usepackage{dcolumn}
\usepackage{bm}

\def\nc{\newcommand}

\def\lsim{\mathrel{\raise.3ex\hbox{$<$\kern-.75em\lower1ex\hbox{$\sim$}}}}
\def\gsim{\mathrel{\raise.3ex\hbox{$>$\kern-.75em\lower1ex\hbox{$\sim$}}}}
\def\etal{{\em et~al.~}}

\def\eg{{\em e.g.~}}

\def\rhs{{r.h.s.~}}
\nc{\half}{\frac{1}{2}}
\nc{\shalf}{\ensuremath{\textstyle \frac{1}{2}}}

\nc{\deldag}{\mathbin{\partial\mkern-10.5mu\big/}}
\nc{\kdag}{\mathbin{k\mkern-10mu\big/}}
\nc{\Pdag}{\mathbin{P\mkern-10mu\big/}}

\nc{\beq} {\begin{equation}}
\nc{\eeq} {\end{equation}}
\nc{\beqa}{\begin{eqnarray}}
\nc{\eeqa}{\end{eqnarray}}

\begin{document}


\title{Spherically symmetric spacetimes in $f(R)$ gravity theories}
\author{Kimmo Kainulainen${}^{1}$}
   \email{Kimmo.Kainulainen@phys.jyu.fi}
\author{Johanna Piilonen${}^{1}$}
   \email{Johanna.Piilonen@phys.jyu.fi}
\author{Vappu Reijonen${}^{2}$}
   \email{Vappu.Reijonen@helsinki.fi}
\author{Daniel Sunhede${}^{1}$}
   \email{Daniel.Sunhede@phys.jyu.fi}
   
\affiliation{${}^{1}$Dept.~of Physics, P.O.~Box 35 (YFL),
   		FIN-40014 University of Jyv\"askyl\"a \\
	${}^{2}$Helsinki Institute of Physics and Dept.~of Physical Sciences,
		P.O.~Box 64, FIN-00014 University of Helsinki, Finland}

\date{\today}

\begin{abstract}

We study both analytically and numerically the gravitational fields of stars in $f(R)$ gravity theories. We derive the generalized Tolman-Oppenheimer-Volkov equations for these theories and show that in {\em metric} $f(R)$ models the Parameterized Post-Newtonian parameter $\gamma_{\rm PPN} = 1/2$ is a robust outcome for a large class of boundary conditions set at the center of the star. This result is also unchanged by introduction of dark matter in the Solar System. We find also a class of solutions with $\gamma_{\rm PPN} \approx 1$ in the metric $f(R)=R-\mu^4/R$ model, but these solutions turn out to be unstable and decay in time. On the other hand, the {\em Palatini} version of the theory is found to satisfy the Solar System constraints. We also consider compact stars in the Palatini formalism, and show that these models are not inconsistent with polytropic equations of state. Finally, we comment on the equivalence between $f(R)$ gravity and scalar-tensor theories and show that many interesting Palatini $f(R)$ gravity models can not be understood as a limiting case of a Jordan-Brans-Dicke theory with $\omega \rightarrow -3/2$.

\end{abstract}

\pacs{04.50.+h, 98.80.-k, 95.35.+d}

\maketitle

%
%

\section{Introduction}
\label{sec:intro}

The observation that the expansion rate of the universe appears to be accelerating~\cite{astier,spergel} has led to a great interest in exploring the possible extensions of the Einstein-Hilbert theory of gravity. A particularly popular class of models involves including nonlinear interactions in the Ricci scalar $R$:
\begin{equation}
	S = \frac{1}{16 \pi G} \int {\rm d}^{4}x \sqrt{-g} f(R) + S_{\rm m}
\label{eq:action}
\end{equation}
where $S_{\rm m}$  is the matter  action. Setting $f(R) = R - 2\Lambda$ corresponds to the canonical Einstein-Hilbert action with a cosmological constant $\Lambda$. Recent interest in these models has followed from the observation~\cite{vollick,carroll} that adding a function $\delta f(R) \equiv f(R)-R = -\mu^4/R$ can give rise to the observed acceleration without a cosmological constant. 

However, it has been proven that when understood as a {\em metric} theory~\footnote{That is, assuming that the affine connection of the spacetime manifold is given by the Levi-Civita connection: $\Gamma^{\rho}_{\mu \nu} \equiv \left\{ {}^{\rho}_{\mu \nu} \right\}$.}, the action (\ref{eq:action}) leads to predictions that are in contrast with the measurements in the Solar System~\cite{chiba,erickcek}. Indeed, as was claimed by Chiba~\cite{chiba}, the Parameterized Post-Newtonian (PPN) parameter $\gamma_{\rm PPN}$ in these theories is $\gamma_{\rm PPN}=1/2$, while the observational constraint requires $\gamma_{\rm PPN}-1 \lsim 10^{-4}$~\cite{obsongamma}. This result has been contested by several authors~\cite{debate,allemandi,ruggiero}, but was recently confirmed by a more direct computation by Erickcek \etal \cite{erickcek}. Both Chiba and Erickcek \etal considered the space-times of stars surrounded by empty vacuum where at large radii $R\rightarrow \sqrt{3}\mu^2\equiv R_{\rm vac}$ for the $\delta f(R)=-\mu^4/R$ model. However, in reality stars and star-systems in galaxies are surrounded by a halo of dark matter and one could ask if this might change the conclusions. 

Indeed, in General Relativity (GR) one finds $R \approx 8\pi G \rho$ (assuming $p\ll \rho$) and this relation becomes the appropriate asymptotic limit for the Ricci scalar in many $f(R)$ models with a constant density background. Typically one assumes that $\rho_{\rm DM} \approx 0.3$ GeV/cm$^3$~\cite{Jungman:1995df}, so that $8\pi G \rho_{\rm DM} \approx 10^6 R_{\rm vac}$. This continuous density field would then appear to have a potentially large influence on the predictions of an $f(R)$ model~\footnote{There is of course also the stellar wind, whose density in the Solar System scales roughly as 5 GeV/cm$^3$(AU/$r_{\odot})^2$, but the detailed form of the density distribution turns out not to be important in what follows.}: for the case $f(R)=R - \mu^4/R$ one has
\begin{equation}
	f(R) = R\; \left( 1 - \frac{\mu^4}{R^2}\right) \, ,
\label{eq:limit}
\end{equation}
and according to the previous argument one expects that $\mu^4/R^2 \approx \mu^2/8\pi G\rho \ll 1$. It would then be tempting to conclude that the theory reduces to GR in a high density environment. If so, then $\gamma_{\rm PPN} \approx 1$ in accordance with GR but in contrast to the vacuum prediction of the theory. This argument was essentially what was used in a recent attempt~\cite{Zhang:2007ne} to show that also metric theories predict $\gamma_{\rm PPN} = 1$ in a dark matter background, in contrast with the vacuum result~\cite{chiba,erickcek}. We show here that introducing a finite dark matter density around a star in fact has no effect on the solutions in metric $f(R)$ gravity.

However, another issue with the metric $f(R)$ theories, which contain derivatives of fourth order, concerns the {\em boundary conditions}. Typically the boundary conditions are set far outside the star for the exterior solutions and to our knowledge no interior solutions have previously been computed in the literature. Our approach here is to set the boundary conditions at the center of the star and compute the solutions throughout, both interior and exterior to the star. We find that the result $\gamma_{\rm PPN} = 1/2$ is robust for a large class of boundary conditions, at least as long as one requires that the metric is finite at the center of the star. These solutions are also unaffected by introduction of dark matter in the Solar System. We also find, however, that another class of more fine tuned solutions exist for which the Ricci scalar settles in the Palatini limit of the trace equation (defined as when derivative terms become negligible), giving $\gamma_{\rm PPN} \approx 1$ even in the metric theories. Unfortunately these solutions are not physically acceptable since they are unstable in time, decaying through the tachyonic instability discovered by Dolgov and Kawasaki~\cite{Dolgov:2003px}. We will argue that our results are not restricted to the $f(R)=R-\mu^4/R$ model, and hence generic metric $f(R)$ theories appear to be ruled out by the PPN data. 

We will also consider finite density effects on stellar solutions in the Palatini formalism. In this case the limit $R \approx 8\pi G \rho_{\rm DM}$ does hold, reducing the theory to GR not only near and inside stellar objects, but essentially at all scales where the matter density is significantly higher than the asymptotic value of the cosmological constant. As a result, Palatini $f(R)$ gravity leads to $\gamma_{\rm PPN} = 1$ to a very high accuracy in the Solar System. These solutions are also stable in time since the Dolgov-Kawasaki instability does not exist in the Palatini formalism~\cite{Sotiriou:2006sf}.

Barausse \etal \cite{Barausse:2007pn} recently claimed that no solution can be found for Palatini $f(R)$ models consistent with a polytropic equation of state (EOS) with index $3/2 < \Gamma < 2$. We do not agree with the interpretation of their results. The problem discovered in Ref.~\cite{Barausse:2007pn} follows from the fact that the Einstein equations in $f(R)$ models depend on $\rho''$ through the  $r$-derivatives of the function $F\equiv \partial f/\partial R$, and that this quantity diverges at the boundary of the star for $3/2 < \Gamma < 2$. However, we will show that this singularity is so weak that, in the case of a neutron star, it becomes relevant only at distances $\Delta r \lsim 0.3$ fermi from the boundary. Defining a sharp boundary to this degree of precision is obviously unphysical and beyond the validity of any polytropic model. Moreover, a polytropic description will always brake down at scales less than the mean collisional distance of the particles in the fluid so there is in fact no physical singularity. As a result, at least all $f(R)$ models where $F \equiv \partial f/\partial R$ is a decreasing function of $R$ should be consistent with the existence of compact GR-like stars. 

Finally, we will consider the scalar-tensor gravity theory (STG) equivalence for metric and Palatini $f(R)$ gravity. The Palatini formalism in particular corresponds to a STG model with no kinetic term in the Einstein frame, and as a result the Palatini field follows the shifting potential extremum with no resistance. This explains why Palatini models can be GR-like, while their metric counterparts fail to obey the PPN constraints. However, Palatini $f(R)$ models whose effective potential in the Einstein frame has negative curvature, $V''_{\rm eff} < 0$, are found to be problematic in the sense that they can not be understood as a limiting case of Jordan-Brans-Dicke theories with $\omega \rightarrow -3/2$.

The paper is organized as follows: section~\ref{sec:metric} introduces a complete set of Tolman-Oppenheimer-Volkov equations for solving the configuration of a static spherically symmetric star in metric $f(R)$ gravity. We derive the Newtonian limit of these equations and compute analytically the predicted value for $\gamma_{\rm PPN}$. We also supplement the analytical work with a direct numerical solution using the complete Tolman-Oppenheimer-Volkov equations with varying boundary conditions and different forms of $f(R)$. In section~\ref{sec:palatini} we consider the Palatini version of the theory and show by both analytic and numerical calculation that $\gamma_{\rm PPN} = 1$ for any Palatini $f(R)$ model where $F\equiv \partial f/\partial R$ is a decreasing function of $R$. We also consider compact Palatini stars with polytropic equations of state in this section. In section~\ref{sec:stg-chameleon} we consider the scalar-tensor equivalence of $f(R)$ theories and section~\ref{sec:summary} contains our conclusions.

%
%

\section{Metric $f(R)$ gravity}
\label{sec:metric}

In the metric formalism one assumes that the Ricci scalar $R$ and the covariant derivative $\nabla_\mu$ are given in terms of the Levi-Civita connection. The field equations are thus obtained by varying the action (\ref{eq:action}) with respect to the metric $g_{\mu \nu}$ only:
\begin{equation}
	F R_{\mu \nu} - \frac{1}{2} f g_{\mu \nu}
                 - \nabla_\mu\nabla_\nu F+ g_{\mu \nu}\Box F
               = 8\pi G T_{\mu \nu} \, ,
\label{eq:eom}
\end{equation}
where $F\equiv \partial f/\partial R$ and $\Box = g^{\mu\nu}\nabla_\mu\nabla_\nu$. In the spherically symmetric case of a static mass distribution, these equations will contain only three independent parameters corresponding to the Ricci curvature $R$ and the $tt$- and $rr$-components of the spherically symmetric metric $g_{\mu\nu}$:
\begin{equation}
	ds^2 \equiv g_{\mu \nu} x^{\mu} x^{\nu} =
		-e^{A(r)}{\rm d}t^2 + e^{B(r)}{\rm d}r^2 + r^2{\rm d}\Omega^2 \, .
\label{eq:metric}
\end{equation}
To solve the functions $R$, $A$ and $B$, it is most convenient to use the $tt$- and $rr$-components of the field equations (\ref{eq:eom}), supplemented by the trace equation for $R$:
\begin{equation}
	\Box F + \frac{1}{3}(FR - 2f) = \frac{8\pi G}{3} T \;,
\label{eq:trace1}
\end{equation}
where $T \equiv T^{\mu}_{\mu} = -\rho + 3p$.  In the static and spherically symmetric case Eqn.~(\ref{eq:trace1}) becomes
\begin{equation}
	F'' + \frac{2}{r}F' + \frac{A'-B'}{2} F' + \frac{e^B}{3}(FR - 2f) 
		= \frac{e^B}{3} 8\pi G T \;.
\label{eq:trace2}
\end{equation}
Note that if the derivatives of $F$ (eventually of $R$) vanish, Eqn.~(\ref{eq:trace1}) reduces to
\begin{equation}
	F R - 2f = 8\pi G T \;,
\label{eq:trace3}
\end{equation}
which for $f(R) = R - \mu^4/R$ has the solution
\begin{equation}
	R_T = \frac{1}{2} \left(-8\pi G T \pm \sqrt{(8\pi G T)^2 + 12\mu^4} 
							\right) \, .
\label{eq:R-trace}
\end{equation}
Given that $p \ll \rho$ and $\mu^2 \ll 8 \pi G \rho$, the solution with a positive sign, corresponding to an asymptotically de Sitter space~\footnote{The solution with a negative sign in the root corresponds to the case with an asymptotically anti de Sitter space~\cite{Kainulainen:2006wz}.}, gives $R \approx 8\pi G\rho$ as mentioned in the introduction. This argument suggests that, in analog to the vacuum limit of Ref.~\cite{erickcek}, the solutions for Eqn.~(\ref{eq:trace2}) should smoothly approach the solutions of (\ref{eq:trace3}) for large enough $r$ in a finite, spatially constant background density.

It is useful to rewrite the field equations (\ref{eq:eom}) in the equivalent form:
\begin{eqnarray}
	G_{\mu \nu} &=& \frac{8\pi G}{F}T_{\mu \nu}
             - \frac{1}{2}g_{\mu \nu} \Big(R - \frac{f}{F} \Big) 
    \nonumber \\
                 & & {} + \frac{1}{F} 
       \left(\nabla_\mu \nabla_\nu- g_{\mu \nu} \Box \right) F \: ,
\label{eq:eom2}
\end{eqnarray}
where $G_{\mu \nu} = R_{\mu \nu} - \frac{1}{2}R g_{\mu \nu}$. The source equations for $A'$ and $B'$ can now be found from the $tt$ and $rr$ components of the field equations~(\ref{eq:eom2}). Using Eqn.~(\ref{eq:trace2}) to get rid of the $F''$ term on the right hand side in the $rr$ equation, one eventually finds:
\begin{eqnarray}
	A' & = & \frac{-1}{1 + \gamma} \left(
			\frac{1 - e^B}{r} - \frac{re^B}{F}8\pi G p
                                   \right.
   \nonumber \\ && \phantom{Hanna} \left.
       + \frac{r e^B}{2}\big( R - \frac{f}{F}\big) + \frac{4\gamma}{r}
		\right) \: , 
\label{eq:sourceA} \\
	B' & = & 
			\frac{1 - e^B}{r} 
			 + \frac{re^B}{F}\frac{8\pi G}{3} (2\rho +  3p)
   \nonumber \\ && \phantom{Hanna}
       + \frac{r e^B}{6}\big( R + \frac{f}{F}\big) - \gamma A' \, ,
\label{eq:sourceB}
\end{eqnarray}
where $\gamma  \equiv  rF'/2F$. Furthermore, from the conservation equation $\nabla_\mu T^{\mu\nu}= 0$ (valid in the Jordan frame~\cite{Koivisto}) one finds the following simple relation between $p'$ and $A'$:
\begin{equation}
	p' = -\frac{A'}{2}(\rho + p) \: .
\label{eq:TOVA}
\end{equation}
When supplemented by an equation of state
\begin{equation}
	p = p(\rho)
\label{eq:eos}
\end{equation}
equations (\ref{eq:trace2}), (\ref{eq:sourceA}-\ref{eq:sourceB}) and (\ref{eq:TOVA}) give a complete generalization of the Tolman-Oppenheimer-Volkov equations for a relativistic, spherically symmetric star in metric $f(R)$ gravity theory.  The above equations, which promote $F$ and $F'$ to be free variables along with $A$ and $B$, appear to be much simpler than the alternative equations derived \eg in Ref.~\cite{Multamaki:2006ym}, where $R$ and eventually $F$ and $F'$  were expressed in terms of $A$ and $B$ and their derivatives.

\subsection{The Newtonian approximation}

As was emphasized by Erickcek \emph{et al.}~\cite{erickcek}, a unique exterior solution for a stellar object is found by matching it with an interior solution in the presence of matter sources. In this section we find such a solution for the Tolman-Oppenheimer-Volkov equations derived in the previous section. If any solution is to be found compatible with the PPN limits, the corresponding metric should be close to that of the Newtonian limit. That is, one can assume that $A, B \ll 1$ and $A',B' \ll 1/r$ in Eqns.~(\ref{eq:trace2}) and (\ref{eq:sourceA}-\ref{eq:sourceB}). 

Another related simplification that we shall make is to neglect the pressure throughout and use fixed density profiles instead. The rationale for this is that we are ultimately only interested in the {\em metric} created by a generic Newtonian matter configuration and not on solving for these configurations self-consistently.  Turning the argument around, Eqns.~(\ref{eq:TOVA}-\ref{eq:eos}) could be used in retrospect to compute the pressure and equation of state that would be needed to create the employed density profile, but since $p \ll \rho$ in the Newtonian limit, that pressure would only give a negligible contribution to the metric and can hence be ignored. In fact, except for implementing $f(R)$ gravity into a full numerical stellar model, our approach may provide the most accurate test for the model, since the equation of state computed in retrospect from a realistic stellar density profile is likely to be more appropriate than simple toy model equations of state (see section (\ref{sub:polytropic}) below).

\subsubsection{The trace equation}

Let us first consider the trace equation (\ref{eq:trace2}) in the Newtonian limit:
\begin{equation}
	F'' + \frac{2}{r}F' + \frac{1}{3}(FR - 2f) 
		\approx - \frac{8\pi G}{3} \rho \;,
\label{eq:trace2_F}
\end{equation}
where we have also assumed that $p \ll \rho$ so that $T \approx -\rho$. Furthermore, define a new variable $d \equiv F - 1$ which parametrizes the deviation from GR. For the particular model under consideration, $f(R) = R-\mu^4/R$, one now obtains the following equation for $d$:
\begin{equation}
	d'' + \frac{2}{r}d' - \frac{\mu^2(1 - 3d)}{3\sqrt{d}} 
                     = - \frac{8\pi G}{3} \rho \, .
\label{eq:trace2_d}
\end{equation}
Note that, save for the nonlinear term, this form is of course independent of the chosen $f(R)$ model. That is, the class of the solutions for $d$ (and hence for $F$ and ultimately for $R$) is completely dictated by the relative size of the nonlinear term to the derivative terms. It is hence instructive to rewrite the nonlinear term and the source in (\ref{eq:trace2_d}) so that their relative size in different environments become more transparent:
\begin{equation}
	d'' + \frac{2}{r}d' = H_0^2 \left(
			\frac{4\Omega_\Lambda(1 - 3d)}{3\sqrt{d}}
			- \frac{\rho}{\rho_{\rm crit}}
		\right) \;,
\label{eq:trace2_d2}
\end{equation}
where $3H_0^2\equiv 8\pi G\rho_{\rm crit}$ is the Hubble expansion and $\sqrt{3}\mu^2 = 4\Lambda$. Assume now that one starts to integrate Eqn.~(\ref{eq:trace2_d2}) outwards from the center of a stellar object with some boundary value $d=d_0$. Obviously, the density ratio $\rho_0/\rho_{\rm crit}$ is enormous at $r=0$ and it completely dominates the evolution unless $d_0$ is fine tuned to zero to a high precision. Indeed, the nonlinear term can influence the evolution of $d$ only at very low densities, where $\rho  \lsim  \rho_\Lambda = \Omega_\Lambda \rho_{\rm crit}$, or at high densities when $d \lsim d_\rho$, where
\begin{equation}
	d_\rho = \frac{\mu^4}{R_\rho^2} \approx  \frac{16\Omega_\Lambda^2}{3}\Big(\frac{\rho_{\rm crit}}{\rho}\Big)^2
\label{compare}
\end{equation}
and $R_\rho$ is given by Eqn.~(\ref{eq:R-trace}) under the assumption that pressure is negligible. At the center of the Sun one has $\rho_0/\rho_{\rm crit} \sim 10^{31}$, and thus the nonlinearity is negligible unless $d \lsim 10^{-62}$!  At the edge of the Sun (defined as the radius that encompasses 99.9\% of the mass) the corresponding value would be $d \lsim 10^{-58}$. Even outside the Sun, within the local dark matter distribution with $\rho_{\rm DM} = 0.3$ GeV/cm$^3$, one finds that the density will dominate over the nonlinear term unless $d \lsim 10^{-12}$.

Barring such fine tuned boundary conditions, the nonlinear term can be neglected in Eqn.~(\ref{eq:trace2_d}), at least during the initial stages of the evolution.  Let us now solve for $d$ in this approximation, setting the boundary conditions at $r=r_0$ with the understanding that in the end $r_0\rightarrow 0$. Integrating Eqn.~(\ref{eq:trace2_d2}) once, one finds:
\begin{equation}
	d'(r) = - \frac{2G\Delta m(r)}{3r^2}- \frac{(r^2d')_0}{r^2} \, ,
\label{eq:dsol1a_0}
\end{equation}
where $(\;)_0$ indicates that an expression is evaluated at $r = r_0$, $\Delta m(r) \equiv m(r) - m(r_0)$, and
\begin{equation}
	m(r) \; \equiv \; \int_0^r {\rm d}r' 4\pi r'^2 \rho \, .
\label{eq:dsol1b}
\end{equation}

Eqn.~(\ref{eq:dsol1a_0}) shows that the boundary value for $d'$ at $r_0 = 0$ plays no role in the solution as long as it is finite, because then $(r^2d)_0' \rightarrow 0$ at $r_0 \rightarrow 0$.  Only considering finite solutions, one can set $r_0 = 0$ and $(r^2d')_0 = 0$ and perform another integration to get:
\begin{equation}
	d(r) = - \int_0^r {\rm d}r' \frac{2G m(r')}{3r'^2} + d_0 \, .
\label{eq:dsol1a}
\end{equation}
Another important observation to be made from Eqns.~(\ref{eq:dsol1a_0}) and (\ref{eq:dsol1a}) is that $d$ gains a {\em negative} contribution from the density evolution. Thus, even if one starts from a value $d_0$, for which the nonlinear term is negligible, the evolution will push $d$ towards the region where nonlinearity does become important.  Whether this happens inside the star or a stellar system depends on the relative size of the matter effect compared to $d_0$, and can easily be estimated in retrospect. For now, assume that $d_0$ is large enough so that nonlinear terms can be neglected throughout.

Outside a star in the Solar System scale, the total mass of the configuration is overwhelmingly dominated by the mass of the star: given $\rho_{\rm DM} = 0.3$ GeV/cm$^3$, the mass of the dark matter component inside a radius $r$ in the Solar System is roughly $M_{\rm DM} \sim 10^{-16}M_\odot$ $(r/{\rm AU})^3 $, where AU is one astronomical unit. This shows that the exterior finite dark matter density has no effect on the solutions, contrary to the claims made in Ref.~\cite{Zhang:2007ne}. Hence, to an excellent approximation, the exterior solution is
\begin{equation}
	d \approx  \frac{2GM}{3r} + d_0  - C \, ,
\label{eq:dsolext1}
\end{equation}
where $M$ is the total mass of the star, $M \equiv m(r_{\odot})$. The constant has been $C$ extracted for convenience so that the $1/r$ law becomes explicit in (\ref{eq:dsolext1}),
\begin{equation}
	C \equiv  \frac{2G}{3} \left( \frac{M}{r_{\odot}}
		+ \int_0^{r_{\odot}} {\rm d}r \; \frac{m(r)}{r^2} \right) \, .
\label{eq:Ceqn}
\end{equation}
Putting in the numbers for a typical star like our Sun, one finds that the scale of the density induced evolution in $d$ is of order $d-d_0 \sim 10^{-5}-10^{-8}$, from the center to distances on the order of AU. This sets the scale for the validity of the approximation of neglecting the nonlinear terms: we conclude that {\em the solution (\ref{eq:dsolext1}) holds for any $d_0 \gsim 10^{-5}$ and for arbitrary finite $d_0'$ at $r_0=0$}. Furthermore, it is not affected by a finite dark matter distribution external to the star. However, for boundary conditions $d_0Ê\lsim 10^{-5}$ a more careful analysis is needed and we will study these solutions in separate in section~\ref{sec:Palatinilimit}.

We have based our argumentation on the idea of integrating the equations  starting from the center of the star with some boundary condition. The Erickcek \etal solution~\cite{erickcek}, on the other hand, was derived with  a boundary condition $R \rightarrow \sqrt{3}\mu^2$ or equivalently $d \rightarrow 1/3$ {\em far away} from the star at $r\rightarrow \infty$.  However, it is obvious from the previous discussion that this boundary condition corresponds to an almost identical boundary condition at the center of the star: $d_0 = 1/3-C \approx 1/3$, up to correction of order $10^{-6}$. This also appears to be the physically best motivated boundary condition, as stars were born by condensation of inhomogeneities from an initially rather smooth matter distribution, whereby one most naturally expects that $R \sim \sqrt{3}\mu^2$ at large scales.

\subsubsection{Source equations for $A$ and $B$}

Let us now turn to the source equations for $A$ and $B$ in the Newtonian limit, armed with the solutions (\ref{eq:dsol1a}) and (\ref{eq:dsolext1}) for $F = 1+ d$. For this solution, $\gamma \equiv rF'/2F \approx rd'/2 = -GM/3r \ll 1$, so that one can expand in this variable as well. Hence, to first order in small quantities:
\begin{eqnarray}
	A' & \approx & \frac{B}{r} - 2d' 
       +\frac{\mu^2 r \sqrt{d}}{1+d} \, , 
\label{eq:sourceA_b} \\
	(rB)' & \approx & 
			  \frac{16\pi G}{3} r^2 \rho 
       + \frac{(\mu r)^2}{3\sqrt{d}(1+d)} \, ,
\label{eq:sourceB_b}
\end{eqnarray}
where we have neglected pressure but kept the nonlinear contribution, rewritten in terms of $d = \mu^4/R^2$. However, given the solution (\ref{eq:dsol1a}) for $d$, it is easy to see that the nonlinear terms are always completely negligible in comparison with the other terms, at least as long as one starts with $d_0 \gsim 10^{-5}$. Neglecting these terms, it is straightforward to obtain the solution
\begin{eqnarray}
	B & \approx & 
			  \frac{4G}{3}\frac{m(r)}{r}\Big|_{r_0}^r + \frac{(rB)_0}{r} \, ,
\label{eq:sourceA_b2} \\
	A & \approx & 
			  \frac{8G}{3}\int_{r_0}^r {\rm d}r' \frac{m(r')}{r'^2} - \frac{(rB)_0}{r}+ B_0 + A_0  \, ,
\label{eq:sourceB_b2}
\end{eqnarray}
where we have kept all integration constants, setting the boundary values at $r=r_0$. In particular, one finds that the exterior solution is:
\begin{eqnarray}
	B & \approx & 
			  \frac{4GM}{3r} + \frac{(rB)_0}{3r} \, ,
\label{eq:sourceA_b3} \\
	A & \approx & 
			  - \frac{8GM}{3r} - \frac{(rB)_0}{3r} + B_0 + 4C  + A_0 \, ,
\label{eq:sourceB_b3}
\end{eqnarray}
where $C$ is given by Eqn.~(\ref{eq:Ceqn})~\footnote{Note that  $A$ becomes canonically normalized in coordinates $t'= 3t/4$ and $r'= 3r/4$.}. Note that the validity of the Newtonian limit for the above solutions can be verified in retrospect from the final forms of the solutions (\ref{eq:dsolext1}) and (\ref{eq:sourceA_b3}-\ref{eq:sourceB_b3}). Apart from arbitrary constants: $A$, $B$, $d \sim GM/r \ll 1$. 

In order to obtain the appropriate asymptotic limit $A \sim 1/r$ for large $r$, one must set $A_0 = -4C - B_0$ in (\ref{eq:sourceB_b3}). Given this constraint one finds in the Newtonian limit:
\begin{equation}
	\gamma_{\rm PPN} \; = \; -\frac{B}{A} 
                 \; = \;  \frac{4GM + 3(rB)_0}{8GM+3(rB)_0} \, .
\label{eq:gamma1}
\end{equation}
Thus, for {\em any finite value of} $B_0$ at the center of the star one obtains 
\begin{equation}
	\gamma_{\rm PPN} \; = \; \frac{1}{2} \, .
\label{eq:gamma2}
\end{equation}
This is of course the well known result by Chiba~\cite{chiba}, which was later confirmed in Ref.~\cite{erickcek} and yet again contested in Ref.~\cite{Zhang:2007ne}. We have shown here that this result holds for a large class of boundary conditions at the center of the star: when $d_0\gsim 10^{-5}$ and for practically any $d_0'$, $A_0$ and $B_0$, as long as one remains in the Newtonian limit and $A \sim 1/r$ for large $r$. Furthermore, this solution is not affected by the dark matter distribution surrounding the star.

Turning the above argument around, one can also see from Eqn.~(\ref{eq:gamma1}) that the condition $\gamma_{\rm PPN} = 1$ exterior to the star necessarily leads to a blow-up in the metric components at the center (again assuming that $d_0 \gsim 10^{-5}$); indeed, $\gamma_{\rm PPN} \approx 1$ would imply $-A_0 \approx B_0 \gg GM/r_0 \rightarrow \infty$ when $r_0 \rightarrow 0$. Such a solution is clearly unphysical, even though the actual divergence of the metric occurs beyond the validity of the Newtonian approximation.
\begin{figure}[!t]  
    \begin{center}
	\includegraphics[width=8cm]{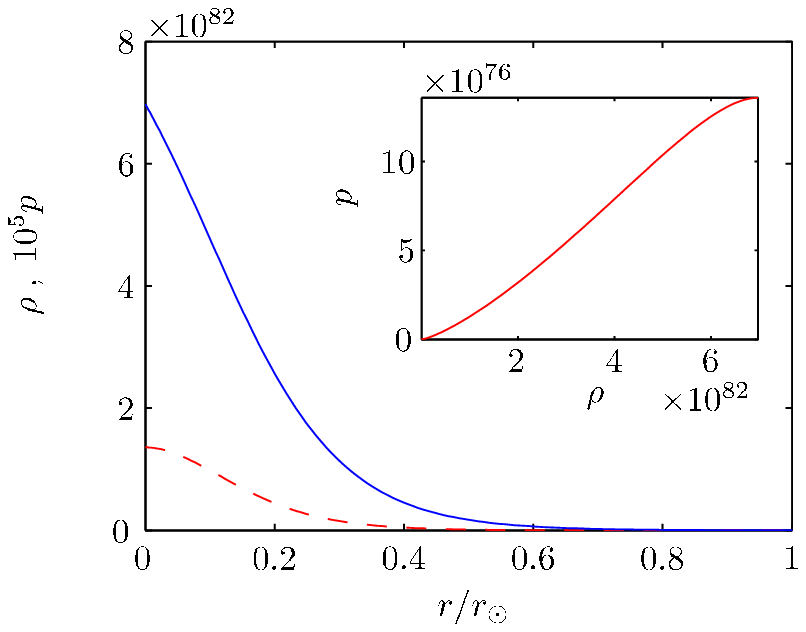}%
    \end{center}
    \caption{Shown is the density profile of the star (solid) and the
    		appropriate pressure needed to support this configuration
    		(dashed) (in units $r_{\odot}^{-4}$). The inset figure shows
    		the equation of state $p=p(\rho)$ corresponding to this solution.}
    \label{fig:profiles}
\end{figure}

\subsection{Numerical analysis}
\label{sec:numerics}

We will in this subsection support our analytic results (which were based on the Newtonian approximation) with numerical solutions of the exact source equations (\ref{eq:sourceA}-\ref{eq:sourceB}) and the trace equation (\ref{eq:trace2}), still using a fixed density profile.  The actual profile used corresponds to the known density profile of the Sun with a central density of $150$ g/cm$^3$ and with a roughly exponential dependence on $r$. The dark matter distribution was taken to be a constant with $\rho_{\rm DM} = 0.3$ GeV/cm$^3$, superimposed on the profile of the star. We display this density profile in Fig.~(\ref{fig:profiles}) (solid line). The dashed line in the figure shows the pressure profile computed in retrospect from equations (\ref{eq:TOVA}) and (\ref{eq:eos}), using a known solution for the metric. The inset figure displays the corresponding equation of state $p=p(\rho)$. These plots show that the assumption $p \ll \rho$ holds, and that our profile corresponds to a reasonable equation of state. 

\begin{figure}[!t]  
    \begin{center}
	\includegraphics[width=8cm]{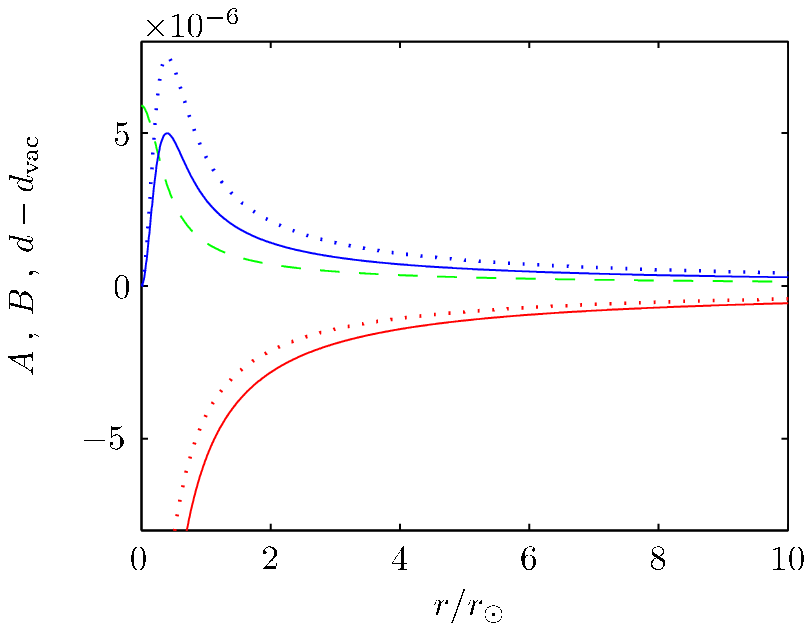}%
    \end{center}
    \caption{Shown are the functions $A$ (red) and $B$ (blue) for the metric
    		$f(R)=R-\mu^4/R$ model (solid) and for GR (dotted).  Also shown is
    		the function $d-d_{\rm vac}$ (dashed green), where $d_{\rm vac}$ is
    		the asymptotic value of $d$ in vacuum.}
    \label{fig:ABd}
\end{figure}

Fig. (\ref{fig:ABd}) shows the evolution of $A$, $B$, and $d-d_{\rm vac}$, where $d_{\rm vac}$ is the asymptotic value of $d$ in vacuum. These plots turn out to be practically independent of the choice of boundary conditions $d_0$, $d_0'$, $B_0$ and $A_0$, as suggested by our Newtonian analysis. (One of course has to set $A_0=-B_0-4C$ in order to ensure the appropriate limit for $A$ at large $r$). The evolution of the corresponding $\gamma_{\rm PPN}$ parameter is shown in Fig.~(\ref{fig:gammaPPN}). Although it of course only makes sense to talk about $\gamma_{\rm PPN}$ for the exterior part of the solution, it is nevertheless interesting to compare how the quantity evolves in GR and in a metric $f(R)$ model, reaching different constant values at the edge of the star. We find that this form of the solution is generic and independent of the boundary conditions as long as $d_0 \gsim 10^{-5}$ and the metric is finite at the center of the star.

\subsubsection{Other forms of $f(R)$}

We have repeated the above numerical analysis for several other models in addition to $f(R) = R - \mu^4/R$.  As suggested by the Newtonian analysis, all $f(R)$ models should give $\gamma_{\rm PPN} = 1/2$, as long as the nonlinear terms can be neglected. We have verified this result by numerical analysis, where have also considered explicitly various other models, including $f(R) = R - \mu^4/R + \alpha R^2$ and $f(R) = R - \beta R^n$, and in particular a model similar to the one suggested in Ref.~\cite{Faulkner:2006ub}: 
\begin{equation}
	f(R) = R + \alpha \sqrt{R} \, ,
\label{tegmarkmodel}
\end{equation}
as a possible candidate for passing the Solar System constraints. We find that all models fail the PPN limit; as long as the model parameters are set to give the correct asymptotic cosmological constant and one does not add a true cosmological constant to the $f(R)$ function, all models produce results that are essentially indistinguishable from the $f(R) = R - \mu^4/R$ model in the Solar System scale~\footnote{Note that also cosmological constraints rule out several of these models~\cite{Amendola}. Nevertheless, local gravity experiments likely puts the most severe constraint on metric $f(R)$ gravity.}.

\subsection{Solutions with $d_0 \lsim 10^{-5}$}
\label{sec:Palatinilimit}

Let us now go back to the class of solutions with very small boundary values for $d_0$. As explained above, in this regime the matter induced evolution of $d$ is strong enough to push the solution to the nonlinear region inside a Sun-like star.  One can argue qualitatively that the resulting solution will be one where $d$ oscillates around the value of $d_\rho$ corresponding to the Palatini limit, Eqn.~(\ref{eq:R-trace}). Indeed, since $R$ is a real number, $d$ must always remain positive in the $f(R) = R - \mu^4/R$ model. However, starting from $d_0 \gg d_\rho$ at $r=r_0$, $d$ will first start to decrease. This evolution is bound to be reversed by the nonlinear term before $d$ becomes negative, but once $d$ starts to increase, the nonlinearity shuts off again and the the finite density effect turns the evolution back towards smaller $d$. As the cycle gets repeated, the result is an oscillatory motion around the Palatini limit, defined  as the solution of the equation (\ref{eq:trace3}). We show an example of this behaviour in Fig.~(\ref{fig:Adosc}). The solid line shows the evolution of $d$, which indeed settles to a damping oscillatory pattern around the Palatini limit, shown by the dotted line. We also display the metric coefficient $A$ (dashed) which, after a short interval of ``Newtonian $f(R)$ evolution'', settles to a converging oscillatory track around a path parallel to the GR solution $A_{\rm GR} - A_0 \approx 0$. The solution for $B$ turns out to be numerically indistinguishable from the corresponding GR solution. In summary, $A$ and $B$ turn out to be very close to the GR solution simply due to the fact that the Palatini solution is virtually indistinguishable from the GR metric (see section \ref{sec:palatini}).

\begin{figure}[!t]  
    \begin{center}
	\includegraphics[width=8cm]{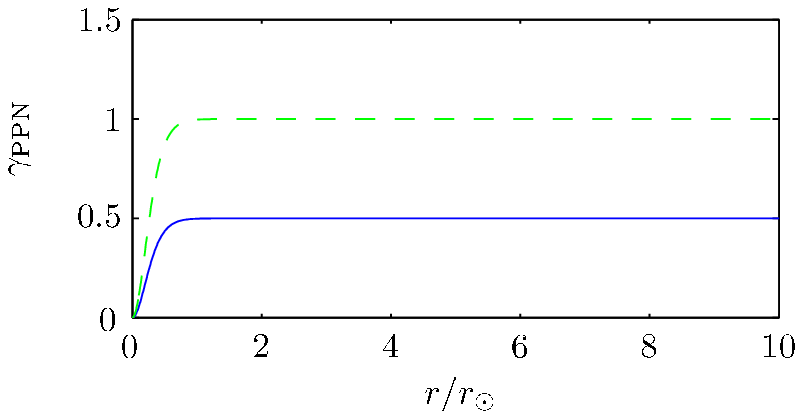}%
    \end{center}
    \caption{Shown is $\gamma_{\rm PPN}$ for a metric $f(R)$ gravity (solid)
    		and the corresponding solution in GR (dashed).}
    \label{fig:gammaPPN}
\end{figure}
Note that at the center of the Sun, the oscillations occur in scale $\sim 10^{-28}r_{\odot}$, so it is not numerically feasible to continue the solution all the way to the surface. We have nevertheless run the code over thousands of oscillation periods, verifing that the solution does indeed stabilize around the Palatini limit. Furthermore, this behaviour is independent of the boundary value $d_0$. (Of course, if one sets $d_0=d_\rho$, the solution will become flat without any oscillations.) The above example used a very small $d_0$, but the qualitative behaviour of the solution should remain the same for any $d_0$ for which the Newtonian evolution is strong enough to bring $d$ to zero inside the star. As a result, it is safe to conclude that for sufficiently small $d_0$ the solution will be such that inside and in particular outside the star $A \approx A_{\rm GR}$ and $B \approx B_{\rm GR}$, so that $\gamma_{\rm PPN} \approx 1$. In practice, the boundary for this result may be somewhat less than $d \lsim 10^{-5}$ since $d$ needs to reach the nonlinear region already close to the center of the star. If not, the initial evolution of $A$ and $B$ will have time to push the metric and eventually $\gamma_{\rm PPN}$ too far from the GR solution.

\subsubsection{The Dolgov-Kawasaki instability}
\label{sec:DKinstability}

\begin{figure}[!t]  
    \begin{center}
	\includegraphics[width=8cm]{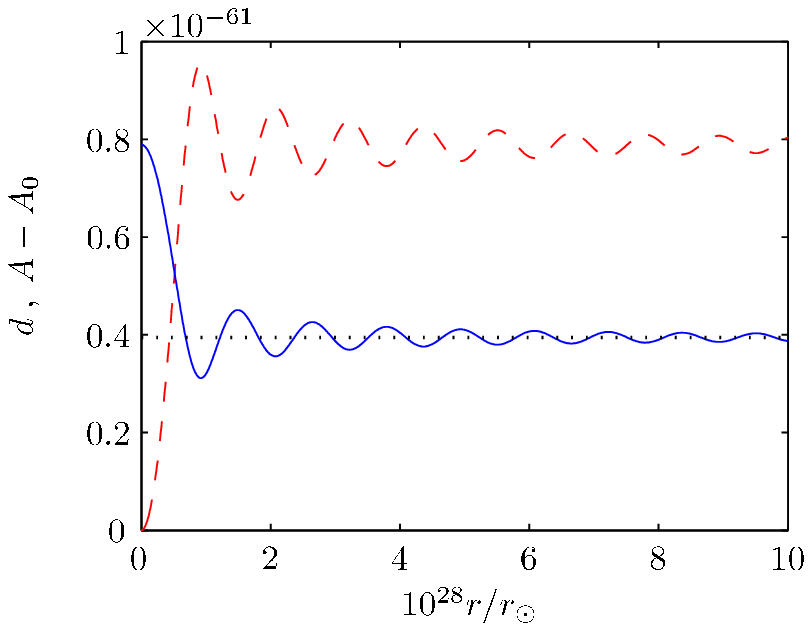}%
    \end{center}
    \caption{Shown are the functions $d$ (solid) and $A-A_0$ (dashed), where
			$A_0$ is chosen such that for GR $A \sim 1/r$ at
			$r \rightarrow \infty$. The dotted line shows the Palatini value
			of $d$, which corresponds to $R \approx 8\pi G \rho$.}
    \label{fig:Adosc}
\end{figure}

The above section explored an attractor solution around the Palatini limit for small values of the boundary value $d_0$. However, it turns out that this class of solutions is related to the well known Dolgov-Kawasaki instability~\cite{Dolgov:2003px} in the $f(R) = R - \mu^4/R$ model. Perturbing around the static solution, $d(r) \rightarrow d(r) + \delta d(t,r)$, and expanding to first order in the perturbation one obtains the following equation:
\begin{equation}
	\partial_t^2 \delta d 
	- \nabla^2\delta d \; \approx \; 
	 \frac{\mu^2}{6d^{3/2}}(1+3d)\delta d  \; \equiv \;  - m_d^2 \delta d\; ,
\label{eq:trace2_deld}
\end{equation}
The negative effective mass squared in this wave equation is what, when restricted to the static limit, gives rise to the damping oscillations in $r$ around the Palatini limit seen in Fig.~(\ref{fig:Adosc}). However, this attractor behaviour in $r$ comes with the price of making the solution unstable in {\em time}. Indeed, expanding the perturbation in Fourier modes, one finds that a mode with wave vector $\vec{k}$ has time dependence (for $m_d^2 < 0$)
\begin{eqnarray}
	\delta d_k(\vec{k},t) \sim e^{\pm i\sqrt{k^2 - |m_d^2|}t} \,,
\label{eq:mode}
\end{eqnarray}
so that there are unstable modes with $k < |m_d|$. This is the instability first found by Dolgov and Kawasaki~\cite{Dolgov:2003px}.

It is important to notice that the magnitude of $d$ controls both the characteristic time scale of the instability, $t_{\rm inst} \sim 1/|m_d|$, and the shortest scale of the unstable modes, $r_{\rm inst} \sim c\,t_{\rm inst}$.  For the GR-like solutions above $d \approx d_{\rho_{\odot}} \approx 4 \times 10^{-62}$ at the center of the Sun and one finds an unfavourable $t_{\rm inst} \approx 8\times 10^{-30}$ sec. Even for $d \approx d_{\rho_{\rm DM}} \approx 3\times 10^{-12}$ corresponding to the dark matter density, the time scale of the decay is still relatively short, $t_{\rm inst} \approx 6$ yr. The Dolgov-Kawasaki instability then clearly forbids the Palatini tracking solutions as physical ones. For solutions with $d \approx d_0 \gsim 10^{-5}$ the instability time is $t_{\rm inst} > 500$ kyr, and it becomes
of order 1 Gyr for $d \approx 1$.  In summary, one can conclude that the metric $f(R) = R -\mu^4/R$ model has both approximately stable solutions with $\gamma_{\rm PPN} = 1/2$ and unstable  solutions with $\gamma_{\rm PPN} = 1$, but that it does not have any sufficiently stable solutions that would also pass the Solar System tests.

Quickly after the discovery of the Dolgov-Kawasaki instability, a way to cure it was suggested in Ref.~\cite{Nojiri:2003ft}. The idea is to add for example a quadratic term $(\alpha/2\mu^2) R^2$ to the theory, after which the effective mass $m_d$ in equation (\ref{eq:trace2_deld}) becomes (for $d \ll 1$):
\begin{equation}
m_d^2 = \frac{\mu^2}{3\alpha - 6d^{3/2}} \,.
\label{eq:alphamass}
\end{equation}
Since the theory is stable if $m_d^2$ is positive, could one stabilize the above GR-like solutions in this way?  A positive thing to this end is that the GR-like solution does remain an attractor for boundary values $d_0 \lsim 10^{-5}$ even in this extended model. However, in order to make the theory acceptable, one has to make sure that the mass is positive everywhere in the Solar System. That is,
\begin{equation}
\alpha > 2d_{\rho_{\rm DM}}^{3/2} \sim 10^{-17} \,.
\end{equation}
Unfortunately, even for this small value of $\alpha$, the function $F$, which controls the density influence on the growth of $A$ and $B$ in Eqns.~(\ref{eq:sourceA}-\ref{eq:sourceB}), becomes enormous inside the star:
\begin{equation}
F_\rho = 1 + \frac{\mu^4}{R_\rho^2} + \frac{\alpha R_\rho}{\mu^2} \approx
  \alpha \frac{8\pi G\rho}{\mu^2} \sim 10^{11} - 10^{14} \,,
\end{equation}
for $\rho \sim \rho_{\odot} \sim 0.1 - 150$ g/cm$^3$. It is obvious that such a value of $F$ would completely shut off the evolution of $A$ and $B$, giving rise to a nearly massless Schwarzschild exterior solution. This argument is quite generic and it would thus seem to be difficult, if not impossible, to stabilize a GR-like solution throughout a stellar system in any metric $f(R)$ model. Note that $\alpha \sim 1$ stabilizes the model at all scales.

%
%

\section{Palatini $f(R)$ gravity}
\label{sec:palatini}

We have recently studied Solar System constraints on Palatini $f(R)$ gravity in Ref.~\cite{Kainulainen:2006wz}, assuming that the star was surrounded by vacuum. This section will include the effects of the nonzero dark matter density to the analysis. As is well known, the trace equation in the Palatini case is an exact algebraic equation identical to (\ref{eq:trace3}):
\begin{equation}
	FR - 2f = 8\pi G T \;,
\label{eq:trace3Pal}
\end{equation}
where the Ricci scalar is a function of both the metric and the independent affine connection, $R = g^{\mu \nu}R_{\mu \nu}(\Gamma)$. As was was first shown in~\cite{Kainulainen:2006wz}, the source equations for $A(r)$ and $B(r)$ become:
\begin{eqnarray}
	A' & = & \frac{-1}{1 + \gamma} \left(
			\frac{1 - e^B}{r} - \frac{e^B}{F}8\pi Grp
			+ \frac{\alpha}{r}
		\right) \: , 
\label{eq:sourceAPal} \\
	B' & = & \frac{1}{1 + \gamma} \left(
			\frac{1 - e^B}{r} + \frac{e^B}{F}8\pi Gr\rho
			+ \frac{\alpha + \beta}{r}
		\right) \: ,
\label{eq:sourceBPal}
\end{eqnarray}
where
\begin{eqnarray}
	\alpha & \equiv & r^2 \left(
			\frac{3}{4}\left(\frac{F'}{F}\right)^2  + \frac{2F'}{rF}
			+ \frac{e^B}{2} \left( R - \frac{f}{F} \right)
		\right) \: , 
\label{eq:alpha} \\
	\beta & \equiv & r^2 \left(
			\frac{F''}{F} - \frac{3}{2}\left(\frac{F'}{F}\right)^2
		\right) \: , 
\label{eq:beta}
\end{eqnarray}
and $\gamma \equiv rF'/2F$ as before. The conservation equation (\ref{eq:TOVA}) relating $p'$ and $A'$ and the equation of state (\ref{eq:eos}) remain unchanged.

Once again, if any solution is to be found compatible with the PPN limits, the corresponding metric should be close to that of the Newtonian limit. That is, one can assume that $A, B \ll 1$ and $\gamma \ll 1$ in Eqns.~(\ref{eq:sourceAPal}-\ref{eq:sourceBPal}). From the trace equation (\ref{eq:trace3Pal}) one knows that for the $f(R) = R - \mu^4/R$ model, $R \approx 8\pi G$ so that also in the Palatini formalism $d \equiv F - 1 \ll 1$ inside any matter distribution for which $8 \pi GT \gg 12\mu^2$. Neglecting pressure, it is then straightforward to show that to first order in small quantities:
\begin{eqnarray}
	A' & \approx & \frac{B}{r} - 2d' \, , 
\label{eq:sourceAPal_b} \\
	(rB)' & \approx & 
			  8\pi G r^2 \rho + (r^2d')' \;  .
\label{eq:sourceBPal_b}
\end{eqnarray}
Except for the terms containing $d$, these equations are identical to the corresponding equations in GR. That is, if the terms containing $d$ are negligible, these models will indeed be compatible with Solar System constraints.

The solution to the above equations is easy to find:
\begin{eqnarray}
	B & \approx & 
			  \frac{2G m(r)}{r}\Big|_{r_0}^r
			  + \frac{(rB)_0}{r}  + rd' - \frac{(r^2 d')_0}{r}\, ,
\label{eq:sourceAPal_b2} \\
	A & \approx & 
			  2G\int_{r_0}^r {\rm d}r' \frac{m(r')}{r'^2}
			  - \frac{(rB)_0}{r} + B_0 + A_0 
 \nonumber\\
      & & {}+ d + \frac{(r^2d')_0}{r} \, ,
\label{eq:sourceBPal_b2}
\end{eqnarray}
where we have kept all integration constants setting the boundary values at $r=r_0$. Now, for the specific model $d = \mu^4/R^2$ it is straightforward to show that for any density profile that is non-singular at the origin, $(r^2d')_0$ will go to zero at $r = 0$. Hence, one finds that the exterior solution is:
\begin{eqnarray}
	B & \approx & 
			  \frac{2GM}{r}
			  + \frac{(rB)_0}{r}  + rd'\, ,
\label{eq:sourceAPal_b3} \\
	A & \approx & 
              - \frac{2GM}{r}
			  - \frac{(rB)_0}{r} + B_0 + 3C + A_0 + d \, ,
\label{eq:sourceBPal_b3}
\end{eqnarray}
where $C$ is given by Eqn.~(\ref{eq:Ceqn}). Furthermore, since $d \sim \mu^4/(8\pi G\rho)^2$ and $d' \sim (\rho'/\rho)d$, these terms are indeed completely negligible compared to $GM/r$. That is, setting $A_0 = -3C-B_0$ in order to obtain the appropriate asymptotic $A \sim 1/r$ for large $r$, one finally obtains:
\begin{equation}
	\gamma_{\rm PPN} \; = \; -\frac{B}{A} 
                 \; = \;  \frac{2GM + 3(rB)_0}{2GM+3(rB)_0} \, ,
\label{eq:gamma1p}
\end{equation}
so that the model is indistinguishable from GR:
\begin{equation}
	\gamma_{\rm PPN} \; = \; 1 \, .
\label{eq:gamma2p}
\end{equation}
\begin{figure}[!t]  
    \begin{center}
	\includegraphics[width=8cm]{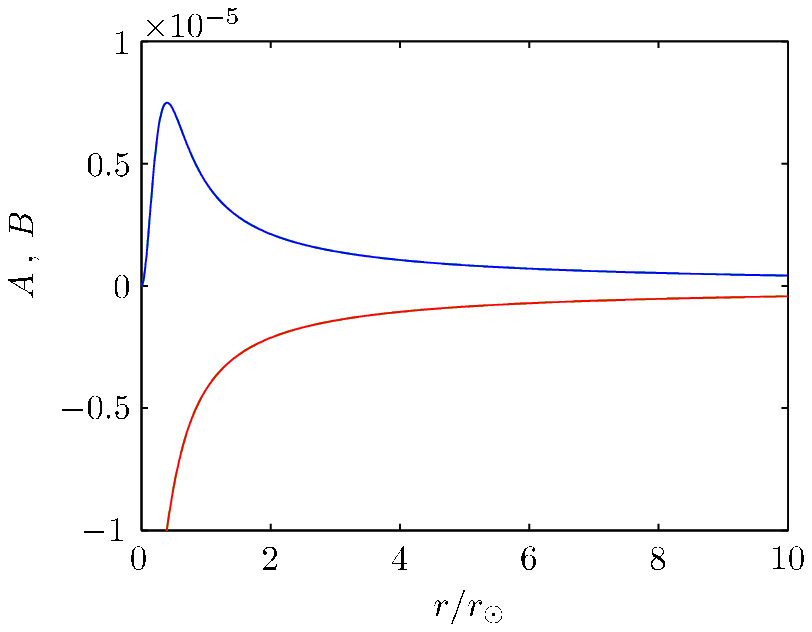}%
    \end{center}
    \caption{Shown are the metric functions $A$ and $B$ for the Palatini
			$f(R)=R-\mu^4/R$ model (solid), which completely overlap with the 
			GR solution (dashed).}
    \label{fig:ABpal}
\end{figure}
The above results are in no way surprising. As was shown in~\cite{Kainulainen:2006wz} (see also~\cite{Sotiriou:2005xe}), for the particular model in question $F$ is extremely close to $1$ inside any density distribution much larger than the effective cosmological constant in vacuum. Hence, $d=F-1$ is completely negligible and the very same supression mechanism brings $F$ to $1$ also in the exterior, dark matter dominated region. That is, in the Palatini formalism, a star surrounded by dark matter will be even more GR-like than a star surrounded by vacuum. Observationally, both scenarios predict a post-Newtonian parameter $\gamma_{\rm PPN}$ indistinguishable from $1$.

To support the analytic results above we have also solved the exact source equations (\ref{eq:sourceAPal}-\ref{eq:sourceBPal}) using the same approach and density profile as in the metric section. Fig.~\ref{fig:ABpal} shows the numerical solution for $A$ and $B$ in the Palatini formalism, which completely overlap with the GR solution. Hence, the curve for $\gamma_{\rm PPN}$ in the Palatini formalism is identical to the GR curve shown in Fig.~\ref{fig:gammaPPN}. As suggested by the Newtonian analysis, these plots turn out to be generic and independent of the given boundary conditions $B_0$ and $A_0$. 

We have also repeated the numerical analysis for several other forms of $f(R)$. We find that for all models where $F \equiv \partial f/\partial R$ is a decreasing function of $R$, $\gamma_{\rm PPN}$ remains indistinguishable from 1 as long as the model parameters are chosen to reproduce the correct asymptotical cosmological constant.

\subsection{Polytropic stars in Palatini $f(R)$ gravity}
\label{sub:polytropic}

A recent paper by Barausse \etal \cite{Barausse:2007pn} considered static, spherical stars with polytropic equations of state, and concluded that Palatini $f(R)$ models would not give physically acceptable alternatives to GR. While the analysis does reveal a genuine qualitative difference between Palatini $f(R)$ gravity and GR, we do not agree with their conclusions. The main criticism of Ref.~\cite{Barausse:2007pn} was based on the argument that for a star with a polytropic equation of state,
\begin{equation}
	p = \kappa \rho_0^\Gamma \, , \qquad
	\rho = \rho_0  + \frac{p}{\Gamma-1} \, ,
\label{eq:polytropic}
\end{equation}
the polytropes with $3/2 < \Gamma < 2$ lead to blow-up of $F''$ and hence also of the metric at the ``edge" of the star, $r = r_{\rm out}$, where $\rho_0 \rightarrow 0$. It is indeed true that due to the $F'$ and $F''$ terms on the \rhs of Eqns.~(\ref{eq:sourceAPal}-\ref{eq:sourceBPal}), the metric may be more
sensitive to rapid density changes in $f(R)$ theories than it is in GR. Since $F''$ shows up in the equation for $B'$, $F''$ blowing up at the boundary leads to a divergence of $B'$ as well. However, we wish to show that this issue has more to do with the peculiarity of polytropic EOS's, when assumed to hold to a mathematically abstract precision at the boundary, than with the theory of Palatini $f(R)$ gravity.

Let us begin by writing
\begin{eqnarray}
   F'  &=& \frac{\partial F}{\partial \rho} \, \rho' \, , \nonumber \\
   F'' &=& \frac{\partial^2 F}{\partial \rho^2} \, (\rho')^2
           + \frac{\partial F}{\partial \rho} \, \rho'' \, .
\label{eq:feqns}
\end{eqnarray}
The partial derivatives $\partial F/\partial \rho$ and $\partial^2 F/\partial \rho^2$ can be computed by using the trace equation (\ref{eq:trace3Pal}) and are in general just some finite numbers at $\rho =  0$.  The behaviour of $F'$ and $F''$ is then completely controlled by the behaviour of $\rho$ at the boundary. Since the continuity equation (\ref{eq:TOVA}) is valid for any $f(R)$ gravity theory in the Jordan frame, one finds that (given a finite and nonzero $A'$) $\rho' = (\partial \rho/\partial p) p' \sim (\partial \rho/\partial p) (\rho + p)$, so that for a polytropic EOS (\ref{eq:polytropic}):
\begin{eqnarray}
   \rho'  &\sim&  \rho^{2-\Gamma} \, , \nonumber \\
   \rho'' &\sim&  \rho^{3-2\Gamma} \, ,
\label{eq:rhoeqs}
\end{eqnarray}
in the limit $\rho \rightarrow 0$, {\em independently of the gravity theory} in consideration. In particular, for the polytropes with $3/2 < \Gamma < 2$, the second derivative $\rho''$ diverges while $\rho$ and $\rho'$ remain continuous in the vacuum limit. Note that the behaviour in Eqn.~(\ref{eq:rhoeqs}) (for $\Gamma > 1$) is only possibleÊif the star has a sharp edge where the density goes to zero at some finite radius $r_{\rm out}$.

It is easy to solve Eqn.~(\ref{eq:rhoeqs}) for the density profile close to the edge:
\begin{equation}
   \rho (r) =  \left\{
      \begin{array}{cc}
         \sim (r_{\rm out}-r)^{\frac{1}{\Gamma -1}} \,, & r \le r_{\rm out} \\
         0 \,, & r > r_{\rm out} 
       \end{array} \right. \,.
\label{eq:rhosolution}
\end{equation}
One hence finds $\rho'' \sim (r_{\rm out}-r)^{(3-2\Gamma)/(\Gamma-1)}\theta(r_{\rm out}-r)$, which indeed blows up at $r=r_{\rm out}$ for $ \Gamma > 3/2$. At first sight, this looks troubling since certain interesting microphysical models predict polytropic EOS's with $\Gamma \in [3/2,2]$. For example, a degenerate and nonrelativistic fermion gas has $\Gamma = 5/3$. However, for this to be a problem, the hard polytropic EOS would have to be valid all the way until $\rho = 0$, which is an unphysical expectation.

First, more realistic EOS's tend to become softer as the density decreases. For example figure~5 in Ref.~\cite{Haensel:2004nu} shows that the effective adiabatic index $\Gamma$, which can be as large as 2.7 at the interior of a neutron star, drops to about 1.25 when $\rho \lsim 10^{10}$ g/cm$^3$. In such a case the metric will be perfectly well-behaved everywhere~\footnote{This behaviour is actually clearly visibile from figure~1 in Ref.~\cite{Barausse:2007pn}.}.

Second, since the function $F''$ is typically controlled by a small parameter (for the $f(R) = R - \mu^4/R$ model, $F''\sim \mu^4$), one would expect that the singularity at the edge should be very weak. To see this quantitatively, let us consider the tidal acceleration close to $r=r_{\rm out}$. For a static, spherically symmetric metric one can show that the equation of geodesic deviation in the radial direction is given by:
\begin{eqnarray}
   \frac{{\rm D}^2 \eta^r}{{\rm D}\tau^2 }
       &=& -\frac{e^{-B}}{4} (2A'' + A'^2- A'B') \,\eta^r \nonumber \\
       &\equiv& - K \eta^r \, ,
\end{eqnarray}
where $\eta^r$ is the radial spatial frame component of the orthogonal connecting vector, $D^2/D\tau^2 \equiv (v^\mu \nabla_\mu)(v^\nu \nabla_\nu)$, $\tau$ is the proper time, and $v^\mu$ is the tangent vector at each point of the timelike geodesic~\footnote{For a more thorough explanation of the involved quantities, see for example Ref.~\cite{d'Inverno:1992rk}}. The angular components are proportional to $e^{-B} A'$ and will hence remain unaffected by the blow-up in $F''$. Using Eqns.~(\ref{eq:sourceAPal}-\ref{eq:beta}) and only keeping the terms proportional to $F''$, one finds the effect of the singularity on the tidal acceleration:
\begin{equation}
   K_{\rm sing} = \frac{e^{-B}}{4}\frac{r^2 F''}{F} \left(
      	3 e^B \lambda + \frac{5}{2} \frac{F'}{rF}
      	+ \frac{3}{2}\left(\frac{F'}{F}\right)^2 \right)
\end{equation}
where $\lambda \equiv \frac{1}{2}(R - f/F)$ is the radially dependent effective cosmological constant.

\begin{figure}[!t]  
    \begin{center}
	\includegraphics[width=8cm]{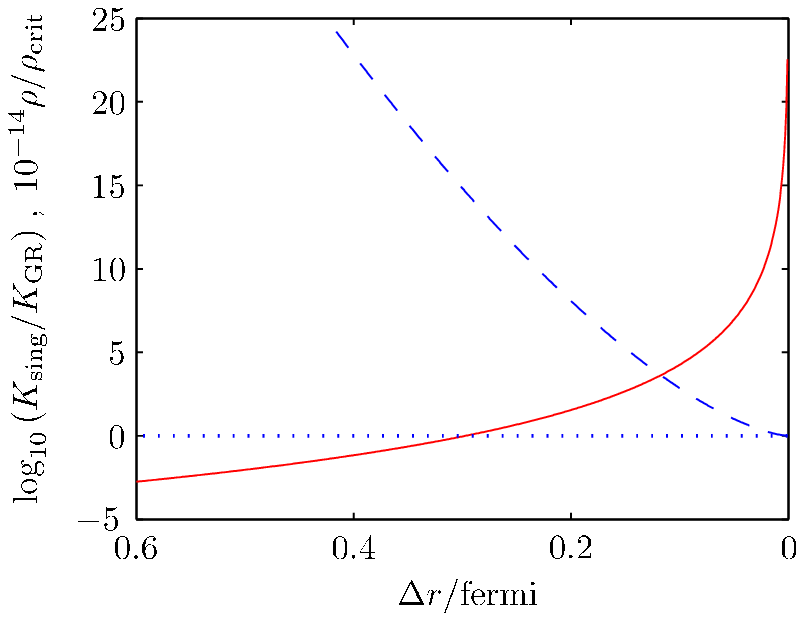}%
    \end{center}
    \caption{Shown is the logarithm of the ratio $K_{\rm sing}/K_{\rm GR}$ for
   		a polytropic star with $\Gamma = 5/3$ in a Palatini $f(R)=R-\mu^4/R$
   		model, as a function of $\Delta r \equiv r_{\rm out} - r$, where
   		$K_{\rm GR} = -2GM/r^3$ is the coefficient of radial tidal acceleration
   		in the Schwarzschild solution. The dashed curve shows the corresponding
   		density in units $10^{14} \rho_{\rm crit}$, which goes to zero at the
   		sharp edge of the star $\Delta r = 0$. We have chosen $M = 2M_{\odot}$
   		and $r_{\rm out} = 10$ km.}
    \label{fig:geodev}
\end{figure}

A reasonable approximation for the magnitude of $A'$ close to the surface is $A' \sim 2GM/r^2$. This will determine the constant in the expression for $\rho$, Eqn.~(\ref{eq:rhosolution}), so that for example for a polytrope with $\Gamma = 5/3$, one obtains:
\begin{equation}
\rho \approx \left( \frac{2GM}{5\kappa r_{\rm out}} \right)^{3/2}
             \left( \frac{r_{\rm out}-r}{r_{\rm out}} \right)^{3/2} \, ,
\end{equation}
where $\kappa = (3/5)(\pi^4/3)^{1/3}m_n^{-8/3} \approx 2.3$ GeV$^{-8/3}$~\cite{Gasiorowicz}. Given this expression for the density, the effect on the tidal acceleration due to the singularity can now be completely determined for a given $f(R)$ model. Fig.~\ref{fig:geodev} shows the size of $K_{\rm sing}$ for $f(R) = R - \mu^4/R$, compared to the size of the radial tidal acceleration in the Schwarzschild solution, $K_{\rm GR} = -2GM/r^3$. For typical neutron star parameters with $M \approx 2M_{\odot}$ and $r_{\rm out} \approx 10$ km, one sees that the tidal acceleration due to the singularity becomes equal to the Schwarzschild value only at a distance $\sim 0.3$ fermi from the surface (which still corresponds to a tiny acceleration $\approx 3\times 10^{-8}$ m/s$^2$ over the distance of one fermi). It should be clear that extending the validity of the polytropic EOS and the mathematical definition of a singular surface to these dimensions is completely unrealistic. We conclude that {\em no generic qualitative} restrictions can be put on Palatini $f(R)$ gravity based on the divergence of $\rho''$ for polytropic EOS's at the boundary.

As was already pointed out in Ref.~\cite{Kainulainen:2006wz}, one does expect that {\em quantitative} constraints should arise from compact stars for the Palatini $f(R)$ models where $F$ is an increasing function of $R$. This expectation seems to be at least partially realized by the results in Ref.~\cite{Haensel:2004nu} on an $f(R) = R + \epsilon R^2$ model. However, the slight deviation from the GR metric found to occur inside the star might not provide a striking enough an observable difference to rule out even these types of models. We plan to return to this issue in a future publication.

Let us finally comment on another recent claim in Ref.~\cite{Olmo:2006zu}, according to which Palatini $f(R)$ gravity would always lead to infinite tidal forces at the boundaries of any objects, such as planets, compact stars, or even atoms. On part, the arguments of Ref.~\cite{Olmo:2006zu} (as was also observed in Ref.~\cite{Barausse:2007pn}) were based on an incorrect identification of the freely falling local inertial frames. As to the behaviour of gravity near sharp boundaries, our arguments above apply. Atoms in particular should rather be thought of as smooth quantum density distributions $\rho_{\rm atom} \sim |\psi_{\rm atom}(x)|^2$, and low energy gravity corrections to such a quantity, either in GR or in a Palatini $f(R)$ gravity, are utterly negligible.

%
%

\section{Scalar-tensor equivalence}
\label{sec:stg-chameleon}

There is a well known equivalence at the classical level between $f(R)$ gravity and Jordan-Brans-Dicke (JBD) scalar-tensor theory (for a review see~\cite{stgequiv}). The general JBD action is given by:
\begin{eqnarray}
	S_\omega &=& \frac{1}{16\pi G}\int {\rm d}x\sqrt{-g} \left(
		\phi R - \frac{\omega}{\phi}(\nabla \phi )^2 - U(\phi) \right)
		\nonumber \\
	&& {}+ S_{\rm m}[g_{\mu\nu},\psi ]\, .
\label{eq:BransDickeJordan}
\end{eqnarray}
The metric $f(R)$ theory is then obtained with $\omega = 0$ and the Palatini theory with $\omega = -3/2$. In both cases the field $\phi$ and its potential $U(\phi)$ are related to the $f(R)$ parameters via
\begin{eqnarray}
   \phi &\equiv& F(R) \quad \Rightarrow \quad R = {\cal R}(\phi) \, ,
   \nonumber \\ 
   U(\phi ) &\equiv& \phi {\cal R}(\phi) - f({\cal R}(\phi)) \, .
\end{eqnarray}
The action (\ref{eq:BransDickeJordan}) is written in the {\em Jordan frame}, defined as the frame where the matter fields couple only to the metric defining the volume element of the theory. That is, the matter action $S_{\rm m}$ is independent of $\phi$. Note that although the field $\phi$ has a ``wrong'' sign for the kinetic term in the Palatini case, it does not mean that the field is unstable. In the Jordan frame, the gravity sector consists of mixing spin-0 and spin-2 fields, and to study stability issues one should move to the Einstein frame.

The Jordan frame is however most convenient for discussing physical observations, because the freely falling local inertial frames correspond to the coordinate systems with locally flat Jordan frame metric. This is so because matter follows the geodesics set by the Jordan frame metric $g_{\mu\nu}$~\cite{Koivisto}. One can in particular show that (see e.g. Ref.~\cite{Esposito-Farese:2000ij}) in a massless Jordan-Brans-Dicke theory the $\gamma_{\rm PPN}$ parameter is simply related to $\omega$:
\begin{equation}
	\gamma_{\rm PPN} = \frac{\omega+1}{\omega+2} \, .
\label{gammaPPNJBD}
\end{equation}
It was through this result that Chiba~\cite{chiba}, using the STG equivalence, first found that metric $f(R)$ gravity give $\gamma_{\rm PPN} = 1/2$.  Turning the argument around, one sees that the PPN bound $|1-\gamma_{\rm PPN}| \lsim 10^{-4}$ puts a strong limit on the Brans-Dicke parameter: $\omega \gsim 10^4$.

Let us now address the PPN bound for the Palatini $f(R)$ gravity in the STG context. Eqn.~(\ref{gammaPPNJBD}) cannot be used as such and the limit $\omega \rightarrow -3/2$ turns out to be very delicate. It is instructive to rewrite the theory in the {\em Einstein frame} by introducing a new metric variable:
\begin{equation}
	h_{\mu\nu} \equiv \phi g_{\mu\nu} \, .
\label{ghrel}
\end{equation}
When written in terms of this metric, the action (\ref{eq:BransDickeJordan}) becomes:
\begin{eqnarray}
	S_\omega &=& \int {\rm d}x\sqrt{-h} \bigg(
		\frac{R_h}{16\pi G} 
			- \frac{2\omega + 3}{6}(\nabla \varphi )^2
		 - V(\varphi) \bigg) 
		 \nonumber \\	&& + S_{\rm m}[\phi^{-1}h_{\mu\nu},\psi]\, .
\label{eq:BransDickeEinstein}
\end{eqnarray}
where $\varphi \equiv \sqrt{3/2}M_{\rm Pl}\log{\phi}$, $V(\varphi ) \equiv  M_{\rm Pl}^2U(\phi )/2\phi^2$, and $M_{\rm Pl} \equiv (8\pi G)^{-1/2}$.
Note that the actions (\ref{eq:BransDickeJordan}) and (\ref{eq:BransDickeEinstein}) are completely equivalent; the conformal scaling of the metric (\ref{ghrel}) is nothing but a convenient change of variables to a frame where the spin-0 and spin-2 degrees of freedom of the gravity sector decouple. There has been a surprising amount of confusion related to this issue, including debates as to which of the two frames should be considered the ``physical" one~\cite{frames}. Of course, neither frame (or any other of an infinity of possible choices for frames) is any more physical than the other; some phenomena are easier to interpret in the Jordan frame while others are best worked out using the Einstein frame. For a clear discussion of these issues see e.g. Refs.~\cite{Damour,Esposito-Farese:2000ij}. For example, the stability of the JBD theory is obscured by the spin mixing in the Jordan frame, but becomes obvious in the Einstein frame, where the kinetic terms in the action (\ref{eq:BransDickeEinstein}) are manifestly positive definite given $\omega \ge -3/2$. For Palatini $f(R)$ gravity, the kinetic term vanishes in the Einstein frame, and so these models correspond to a boundary between stable and unstable JBD models.

\begin{figure}[!t]  
    \begin{center}
    \includegraphics[width=8cm]{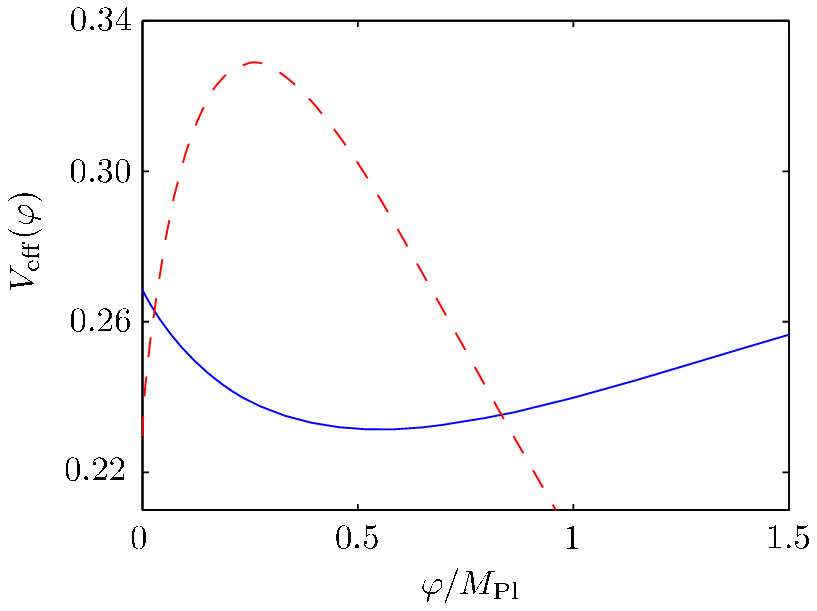}%
    \end{center}
    \caption{Shown is the effective potential $V_{\rm eff}(\varphi)$ (in units
            $\mu^2M_{\rm Pl}^2$) as a function of
            $\varphi/M_{\rm Pl}\equiv \sqrt{3/2}\log(F/F_{\rm min})$ for the
    		model $f(R) = R - \mu^4/R + (\alpha/2\mu^2) R^2$. The solid line
    		corresponds to $\alpha = 0.8$ with $\rho = 5\rho_{\rm crit}$ and the
    		dashed line to $\alpha = 0.01$ with $\rho = 0.1\rho_{\rm crit}$. }
    \label{fig:potential}
\end{figure}
The equation of motion for $\varphi$ in a generic JBD theory follows from the action (\ref{eq:BransDickeEinstein}):
\begin{equation}
   \frac{2\omega + 3}{3}\Box \varphi - V'(\varphi )
   + \frac{\rho }{\sqrt{6}M_{\rm Pl}}e^{-\sqrt{8/3}\varphi/M_{\rm Pl}} = 0 \,,
\label{eq:eomforvarphi}
\end{equation}
where we have neglected pressure, and the exponential factor appears since $\rho$ is the Jordan frame energy density. Now, for Palatini $f(R)$ gravity the coefficient in front of the kinetic term is strictly zero, and the resulting equation
\begin{equation}
   V'(\varphi )
      - \frac{\rho }{\sqrt{6}M_{\rm Pl}}e^{-\sqrt{8/3}\varphi/M_{\rm Pl}}
   \equiv V_{\rm eff}' = 0 \,,
\label{eq:eomforvarphi2}
\end{equation}
is easily seen to be equivalent with the trace equation (\ref{eq:trace3Pal}). The Palatini field $\varphi$ is then completely fixed by the potential $V(\varphi)$ (i.e. the form of the function $f(R)$) and the matter density: the field is constrained to sit at the {\em extremum} of $V_{\rm eff}$. As is seen in Fig.~(\ref{fig:potential}), this extremum can correspond to either a minimum or a maximum of the potential, depending on the model parameters. For the model $f(R)=R-\mu^4/R$ in particular, this extremum is a maximum. In the metric version of the $f(R)$ theory ($\omega = 0$), the potential remains the same, but the field is no longer constrained to stay at the extremum value. This is obviously the origin of the Dolgov-Kawasaki instability in the STG language and it affects all models with $V''_{\rm eff} < 0$.

Now, consider a JBD theory in the limit $\omega \rightarrow -3/2$. In this limit the solution for the trace equation (\ref{eq:eomforvarphi2}) becomes an infinitely strong attractor in the class of static solutions for any model with $V''_{\rm eff} < 0$. However, at the same time this solution becomes {\em arbitrarily} unstable in time. Indeed, perturbing the field $\phi$ around the static solution, $\phi(r) \rightarrow \phi(r) + \delta \phi(t,r)$, it is straightforward to show that a Fourier mode $\delta \phi_k$ of the linearized perturbation obeys the equation
\begin{equation}
   \partial_t^2 \delta \phi_k \approx
      \left( k^2 + \frac{3}{2\omega+3}V''_{\rm eff} \right) \delta \phi_k \,.
\end{equation}
In the limit $\omega \rightarrow -3/2$, {\em all} modes become unstable at the same time as their characteristic decay time goes to zero for $V''_{\rm eff}<0$. This argument shows that it is not possible to obtain a Palatini theory with an effective potential $V''_{\rm eff} < 0$, as a limiting case of a JBD theory with $\omega \rightarrow -3/2$. This is true in particular for the $f(R)=R-\mu^4/R$ model.  For models with $V''_{\rm eff} > 0$, the continuous limit does exist however. Nevertheless, a theory with sufficient stability is not necessarily acceptable as a Palatini $f(R)$ gravity model. For example, stabilizing the vacuum of a JBD model with a potential derived from the model $f(R) = R-\mu^4/R + (\alpha/2\mu^2)R^2$, sets the bound $\alpha \gsim 1$. However, we have seen that in the Palatini limit this leads to an enormous value of the function $F$ (see section \ref{sec:DKinstability}), giving rise to an unacceptable exterior metric.  Even though Palatini $f(R)$ theories are obtained, and can be considered independently of JBD theories, the fact that a smooth limit does not exist for $V''_{\rm eff} < 0$ models is troubling and may indicate some fundamental problem with such theories.

%
%

\section{Conclusions}
\label{sec:summary}

In this paper we have studied the interior and exterior spacetimes of stars in $f(R)$ gravity theories. We started by deriving the generalized Tolman-Oppenheimer-Volkoff equations for spherical hydrostatic equilibrium, for both metric and Palatini versions of the theory. These equations were solved analytically in the Newtonian limit, and the results were supplemented with numerical calculations. Our analysis showed that metric $f(R)$ theories, for the major part of the parameter space, predict a PPN parameter $\gamma_{\rm PPN} = 1/2$ in contrast with Solar System observations. These solutions were independent of finite matter configurations (dark matter) surrounding the star. 

We also found a class of solutions corresponding to boundary values where $F_0-1 \lsim 10^{-5}$, for which $\gamma_{\rm PPN} = 1$ in metric $f(R)$ models. However, these solution are unstable against decay in time and are hence unphysical. Our analytic results were derived mainly in the context of the specific $f(R) = R -\mu^4/R$ model, but our numerical analysis supports the conclusion that the metric $f(R)$ models (beyond the Einstein-Hilbert limit $f(R) = R - 2\Lambda$) either predict $\gamma_{\rm PPN} = 1/2$, or are unstable. However, the Palatini versions of these theories were found to be easily compatible with the Solar System constraints. In addition, we showed that compact stars with polytropic EOS's are consistent with Palatini $f(R)$ gravity. Thus there appears to be no support for a recent claim for a ``no-go theorem'' for Palatini $f(R)$ gravity, based on the use of a polytropic EOS~\cite{Barausse:2007pn}.  

We finally considered the equivalence between $f(R)$ gravity and scalar-tensor theory. While both the metric and the Palatini versions of the theory can formally be viewed as a particular case of Jordan-Brans-Dicke theory, understanding Palatini $f(R)$ gravity as a {\em limiting} case of a Jordan-Brans-Dicke theory with $\omega \rightarrow -3/2$, is not possible at least for models with $V''_{\rm eff} < 0$. This is somewhat troubling and may indicate some fundamental problem with such theories.

%
%

\begin{acknowledgments}
We thank Tuomas Multam\"aki and Iiro Vilja for useful discussions. This work was partially supported by a grant from the Emil Aaltonen Foundation (JP), the Magnus Ehrnrooth Foundation, the Academy of Finland grant 114419 (VR), and by the Finnish Cultural Foundation (DS). Finally, we also acknowledge the Marie Curie Research Training Network HPRN-CT-2006-035863.
\end{acknowledgments}

%
%

\end{document}